# IRVO: an Interaction Model for Designing Collaborative Mixed Reality Systems

*René Chalon & Bertrand T. David*

ICTT - Ecole Centrale de Lyon
36, avenue Guy de Collongue, 69134 Ecully Cedex, FRANCE
Rene.Chalon@ec-lyon.fr, Bertrand.David@ec-lyon.fr

## Abstract

This paper presents an interaction model adapted to mixed reality environments known as IRVO (Interacting with Real and Virtual Objects). IRVO aims at modeling the interaction between one or more users and the Mixed Reality system by representing explicitly the objects and tools involved and their relationship. IRVO covers the design phase of the life cycle and models the intended use of the system. In a first part, we present a brief review of related HCI models. The second part is devoted to the IRVO model, its notation and some examples. In the third part, we present how IRVO is used for designing applications and in particular we show how this model can be integrated in a Model-Based Approach (CoCSys) which is currently designed at our lab.

## 1 Introduction

In the HCI community, Mixed Reality is an important research area for interactive system designers. Mixed Reality systems can be defined as systems which mix real and virtual objects in a coherent way in order to create new tools, close to usual objects as perceived and used by users, but with specific capabilities provided by the computer system with as few technological constraints as possible (MacKay, 1998). According to (Milgram & Kishino, 1994), Mixed Reality is a continuum between reality and virtuality with intermediate steps: augmented reality (AR) and augmented virtuality (AV). Over the last decade, many applications and technologies have been developed in this field. Most of these applications are home made and were developed in an independent and autonomous way. Therefore, it is important to propose a comprehensive model and an associated design methodology.

In a first part, we present a brief review of existing models of Human-Computer Interaction (HCI) which consider real artifacts used in the interaction between the user and the system and we outline their limits regarding Mixed Reality modeling. In a second part, we present our IRVO model and its notation in more detail and illustrate it with two examples. In a third part, we show how IRVO can be integrated in a Model-Based Approach, CoCSys, we are currently designing at our lab. We only focus in this paper on the IRVO used for designing mixed reality software. The relationship between IRVO and the software architecture modeling of the Mixed Reality system is not included in the scope of this paper but is outlined in the conclusion.

## 2 HCI Models and real artifacts

HCI models fall into two main categories, interaction models and architectural models (Beaudouin-Lafon, 2000): an interaction model is a set of principles, rules and properties that guide the design of an interface, whereas an architectural model describes the functional elements in implementation of the interface and their relationships. Because we are focusing on the design of mixed reality systems from a HCI outlook, we only consider interaction models in this paper.

For WIMP (Windows, Icon, Menus and Pointing devices) user interfaces, many HCI models have been developed but very few consider the real artifacts contributing to the interaction: they almost all assume the mouse/keyboard for input and the screen for output and then are limited for modeling more ‘exotic’ devices. Only a few models go beyond this limit and fewer still consider mixed reality explicitly.

Beaudouin-Lafon proposed an interaction model for Post-WIMP interfaces that he called “**Instrumental Interaction**” (Beaudouin-Lafon, 2000). The objects of the task are called Domain objects and are manipulated with interaction instruments which are computer artifacts made up of two parts: the device used to manipulate the interface and the digital part which is the representation on the screen. These interaction instruments are two-way transducers between the user and the domain objects: users act on the instrument, which transforms the user's actions into commands. Users control their actions through the reactions of the instruments and manipulated objects. The instruments also provide feedback as the command is carried out on target objects. This model is aimed at computer interface design but does not at present consider explicitly mixed reality. This model formed the starting point for our IRVO model, in particular for modeling the tools.

In the field of tangible interfaces (which could be seen as a subset of augmented virtuality), Ullmer and Ishii propose a model, called **MCRpd**, standing for model-control-representation (physical and digital) (Ullmer & Ishii, 2000). This model is based on MVC (Burbeck, 1992): M and C are Model and Control, as in MVC, and View is divided into "physical representations" (rep-p) for the physically embodied elements of tangible interfaces and "digital representations" (rep-d) for the computationally mediated components without embodied physical form. This model is limited to tangible interfaces. Although this model is based on MVC, which is an architectural model (and even a framework !), this model is very conceptual and few elements are given to make it operational.

The **ASUR** model is proposed by (Dubois, Nigay & Troccaz, 2001) as a model dedicated to Augmented reality: the user (U) interacts in the real world with the object of the task (Rtask) through a tool (Rtool); the computer system (S) through input adapters (Ain) and output adapters (Aout) can augment either the action of the user or the perception or both. Adapters and real objects are characterized by:

- the human sense involved in perceiving data from such components,
- the location where the user has to focus,
- the ability to share data among several users.

According to Dubois, ASUR has several limits. These include:

- Virtual tools and objects are not represented by ASUR and therefore it cannot model Augmented Virtuality applications properly. This limit is now solved by a recent extension called ASUR 2004 (Dubois et al., 2004) which adds new components (Stool, Sobject and Sinfo) by opening the S 'black box' of ASUR.
- Only one user is represented: ASUR cannot model collaborative applications.

These limits were the main starting point for our IRVO model.

Renevier proposed recently a new notation for mobile mixed collaborative systems (Renevier, 2004). This notation aims at describing scenarios of use in a graphic way instead of the classical textual approach. This notation is almost complete but is not really suited for the design phase of the application: to get round this drawback, Renevier suggests switching to the ASUR notation at this stage.

The Model-Based Approach (MBA) is another approach for user interface development. In this field, (Trevisan, Vanderdonckt & Macq, 2003) applied MBA to Mixed Reality systems. They propose a set of different models to cover all the requirements of an AR system:

- User model which represents user characteristics and their roles. For MR purposes, user location can be added if necessary to this model.
- Task model is a classical representation of the tasks that users need to perform with the application, generally modeled as a tasks tree.
- Domain model which represents the data and the operation (objects) that the application supports; in mixed reality these objects can be either real or virtual objects.
- Presentation model which represents the structure and content of the interface. In MR applications, the spatial and temporal integration of virtual data and real objects must be taken into account for augmented objects.
- Dialog model which describes dynamical aspects of the interface.
- Application and platform model which represents hardware and physical devices. In mixed reality systems, physical devices could be numerous and their characteristics (resolution, accuracy, etc.) are of prime importance.

We also propose to use a Model-Based approach combined with our IRVO model in order to integrate the model in a more generic design process.

## 3 IRVO Model

Starting from the limits of existing HCI models for modeling MR applications, we designed our own model, IRVO (Interacting with Real and Virtual Objects). In particular, the limits of the ASUR model (identified in the previous section) were one of our starting points and therefore IRVO is very similar to ASUR in many aspects.

IRVO aims at modeling the interaction between one or more users and the Mixed Reality system by representing explicitly the objects and tools involved and their relationship. IRVO covers the design phase of the life cycle and models the intended usage of the system. It does not model the software architecture and therefore does not cover the realization phase, but the link between the two phases has been studied (Chalon & David 2004), (Chalon, 2004).

### 3.1 Main entities and relationships

In IRVO we represent 3 main categories of entities (Figure 1a):

- The **user** (U) or more generally users in a collaborative system.
- The objects which can be perceived or manipulated by users. They are either **domain objects** (O) on which the user is focusing for achieving his/her task, or the **tools** (T) which are intermediate objects for helping the user act on domain objects.
- The **internal model** (M) of the application which represents the computer application without the concrete presentation layer.

If the user is clearly in the real world and the application model inside the virtual world, in the case of mixed reality, both tools and objects could either be real or virtual. Figure 1b shows the 4 possible tool-object relationship cases between real (Tr) or virtual (Tv) tools and real (Or) or virtual (Ov) objects. These cases are not exclusive because some objects can be 'mixed', i.e. are composed of a real part and a virtual part. According to the definition of (Milgram & Kishino, 1994), real objects are any objects that have an actual objective existence and can be perceived directly whereas virtual objects are objects that exist in essence or effect, but not formally or actually and must be simulated to be viewed.

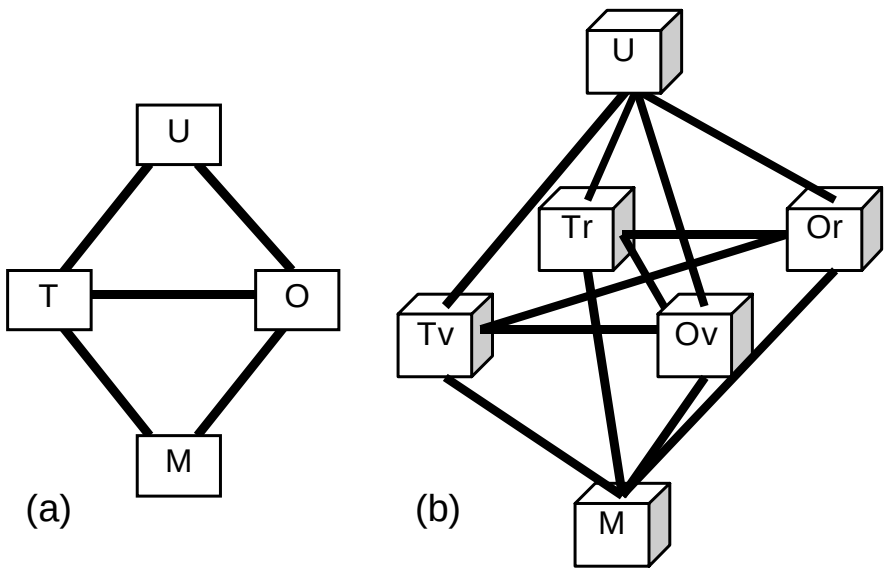

**Figure 1:** IRVO model: the main entities and their relations.

The transmission of information between real and virtual worlds takes place via special IRVO entities called *transducers* which are the only entities in IRVO that are allowed to straddle the R/V boundary.

### 3.2 IRVO notation

#### *3.2.1 Boundaries*

Boundaries are not entities but a means to represent some properties of the entities. There are 2 kinds of boundaries:

- Between the real and virtual world (Figure 2, item ①), represented as a horizontal dashed line, which can be crossed by relationships with the help of transducers.
- Between different places in the real world (Figure 2, item②), represented as a vertical plain line, to show the "opacity" of this boundary, i.e. no relations are allowed to cross this boundary. However, if two places are next to each other, sounds can cross the walls (for example people could communicate by voice without seeing them) and this is represented as item ③ in Figure 2. Alternatively we can represent half-silvered mirrors as in item ④ in Figure 2 (place 4 can view place 3 but place 3 cannot view place 4).

We consider there is no physical border in the virtual world when networking technologies produce an "ubiquitous" environment or CyberSpace. Of course, this is not fully transparent to users: for example, transmission delays can be perceived and can even be disturbing for real time audio communications.

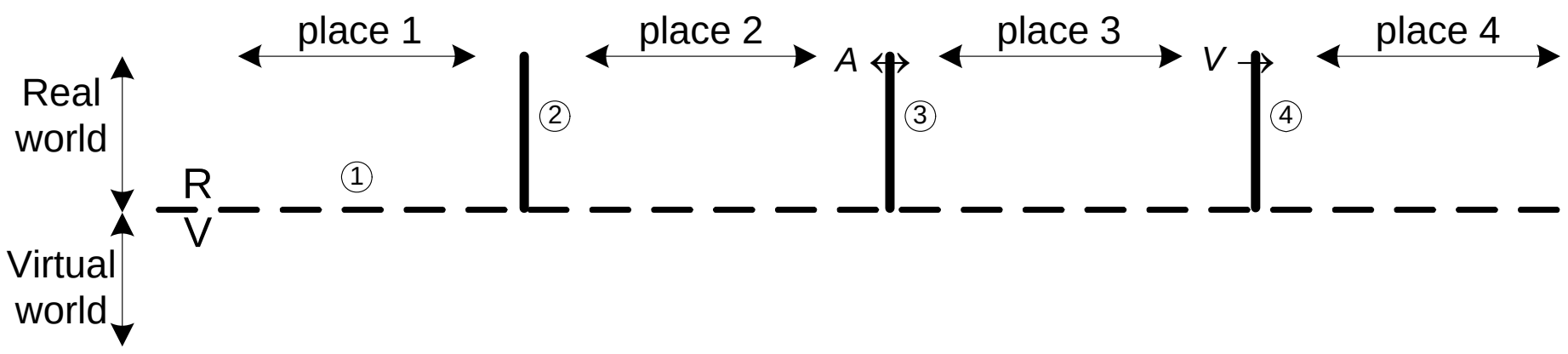

**Figure 2**: Boundaries representation.

### *3.2.2 Entities*

**User** (U) is represented mainly by channels which s/he can use as in Figure 3a; we consider:

- The visual channel (V): mainly as input (eyesight) and as output (direction of sight used by eye-tracking devices).
- The audio channel (A): as input (hearing) and as output (voice: talking, singing, etc.).
- The kinesthetic/haptic channel (KH): as output (handling, grasping, gesture, etc.) and as input (sense of touch).

Other senses, such as taste (T) and sense of smell (S), could also be considered if necessary and easily added to the representation.

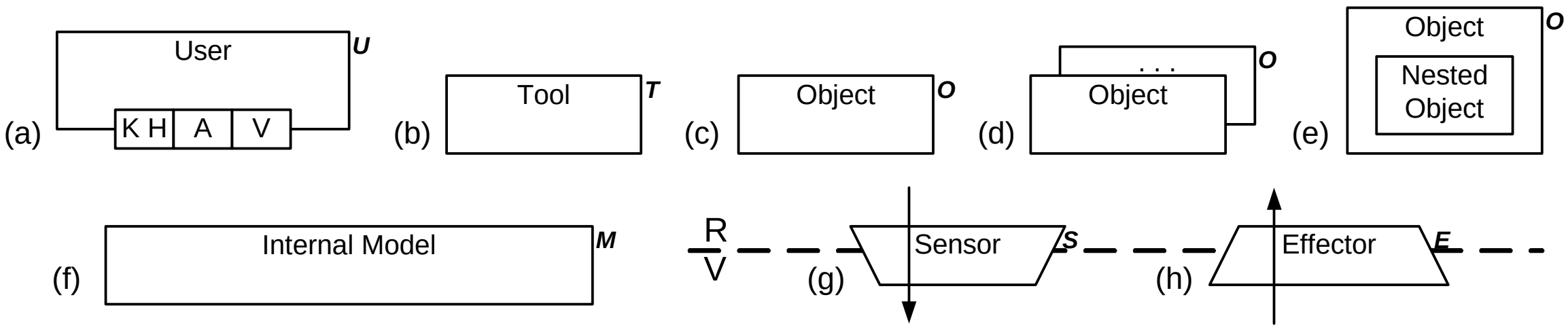

**Figure 3**: IRVO entities.

**Objects** are represented in Figure 3b and Figure 3c: the distinction between **tools** and **domain objects** is only achieved by the "tag" (T or O) placed in the top right-hand part of the rectangle. These tags can also show that the object is real or virtual (Or, Ov, Tr, Tv), which is redundant with its position in relation to the R/V boundary. Stacks (Figure 3d) are used to show a collection of objects of the same kind and nested notation (Figure 3e) to show that some objects are sub-parts of other ones. The **Internal model** (M) is represented like objects with the tag M (Figure 3f). It represents the software behavioral model: it maintains consistency between artifacts and is responsible for object augmentation.

To communicate between real and virtual worlds, information from the real world has to be transformed into digital data, which is carried out by **sensors** (Figure 3g); the reverse operation is performed by devices known as **actuators** or **effectors** (Figure 3h). As a rule, we call them **transducers** by analogy with physical transducers. Because transducers only transform the nature of information (real to virtual world or back), they do not participate directly in the interaction and it is not mandatory to represent them on diagrams. In IRVO, transducers model the functionality of translating real phenomena into virtual data or *vice versa*, and are distinguished from devices. For example, some complex devices can provide several functionalities and therefore are represented by two or more transducers in IRVO models.

### *3.2.3 Relationships*

Relationships (graphically represented by arrows) represent the exchange of information between entities. A relationship could represent an **action** (arrow coming from a user U) or a **perception** (arrow ending at a user) of tools, objects, and more generally, environment. The user channel (KH, A or V) where the arrow starts or ends gives the nature of the information. A relationship could also represent **communication** between two users. Between tools and objects, relationships represent the **action** of tools over objects. A relationship represented as a dashed line means this relationship exists but is not important regarding the current task (for example the haptic feedback of a mouse is not important for moving it left/right or up/down, because the feedback is mainly provided by seeing the pointer on the screen).

Transducers are crossed by relationships to show they convert the nature of information (real world to virtual world) but not the meaning and therefore these transducers do not directly participate in the interaction loop. For example the 2D movements of a pen over the pages are translated into x and y coordinates in the virtual world to be accepted as input of a virtual object. Relationships could be more precisely characterized by using the multimodal properties of (Bernsen, 1994) as in ASUR (Dubois et al., 2001).

### *3.2.4 Representation of Mixed objects*

In the IRVO model, objects cannot cross boundaries: ‘**mixed objects’** are modeled as a composition between a real object (Or) and a virtual object (Ov). This composition is represented by a rectangle in dashed lines encompassing the real and virtual objects (Figure 4). To express the fact that a user perceives the mixed object as a whole, the perception of real and virtual parts is merged with the ⊕ operator.

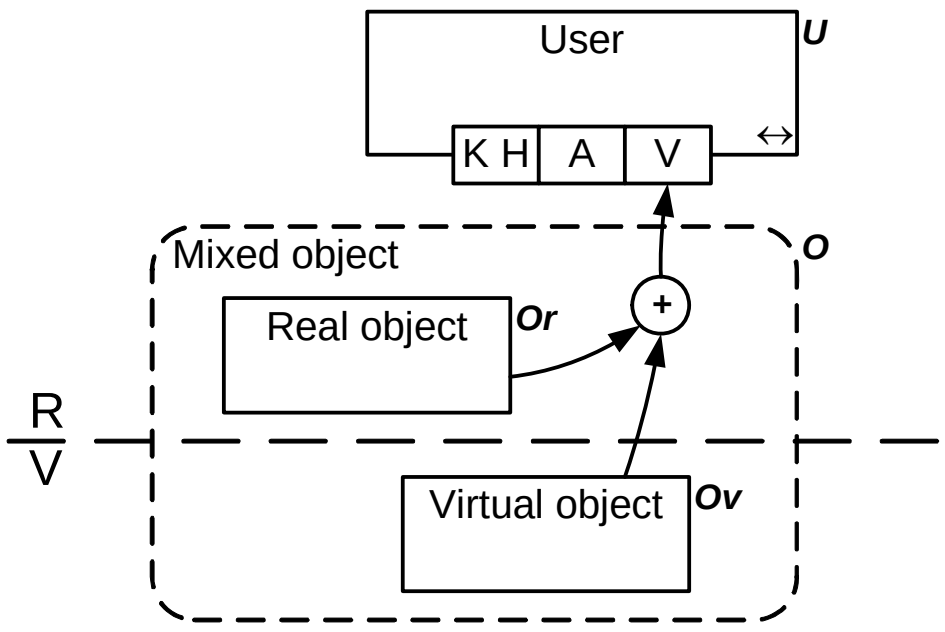

**Figure 4:** Mixed object represented as a composition of objects.

### *3.2.5 Mobility of entities*

The entities (excepted M) can be fixed or mobile within the environment. This property is represented by a symbol in the bottom right-hand part of the entity. To represent the fact that an entity can move, the ‘↔’ sign is used (Figure 5a). If the entity cannot move during the task, a ‘×’ symbol is used (Figure 5b) and if the entity cannot move at all (i.e. during all the tasks of the application), ‘⊗’ the sign is used (Figure 5c). These symbols are just a summary: the exact nature of the mobility property should be specified outside the diagram.

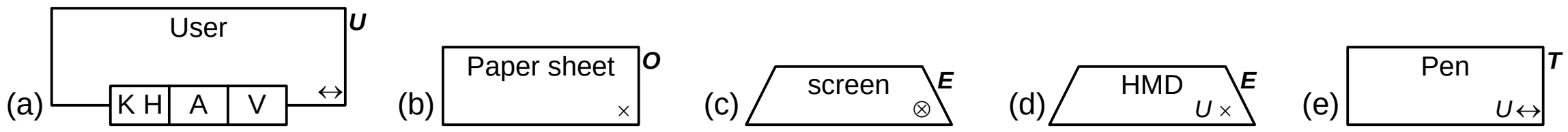

**Figure 5:** representation of the mobility property.

If these signs are preceded by the name of another entity, then the mobility property is not absolute but relative to this entity: for example in Figure 5d, an HMD (Head Mounted display’) is worn by the user (‘*U*×’) and then moves with him. In Figure 5e, Pen (T) is linked to User (U), because it is held by him/her, but the link is not rigid and therefore the notation ‘*U*↔’ is used.

The mobility symbol is not mandatory: if the symbol is not present it leaves the mobility property unspecified, excepted for nested objects which are assumed to be linked (‘×’) by default to the embedding object.

## 3.3 Examples

We applied the IRVO model to 44 mixed reality applications taken from the literature with success. We present here as examples, the modeling of two well-known Augmented Reality applications: the Audio Notebook (Stifelman, 1996) and the DoubleDigitalDesk (Wellner, 1993).

### *3.3.1 Audio Notebook*

The Audio Notebook was proposed by (Stifelman, 1996). This application allows a user to capture and access an audio recording of a lecture or meeting in conjunction with notes written on paper. The audio recording is synchronized when the user turns pages and when s/he points with the pen to a location in the notes. The IRVO model of this application for the browsing task is represented in Figure 6. The paper notebook is the real object of the task (Or) and is composed of several pages (represented as a stack). The page currently viewed is detected by an appropriate sensor (S) which starts reading the audio record represented as a virtual object (Ov). This record is heard, thanks to the speaker (modeled as an effector E), simultaneously with reading of the page. Therefore the Audio Notebook appears to be an ‘augmented object’ (represented by the dashed line tagged ‘O’) made of the real notebook and the virtual audio augmentation. The operator ⊕ is not used as in Figure 4 because perception uses two different human senses (visual and audio): this is a case of multimodal perception.

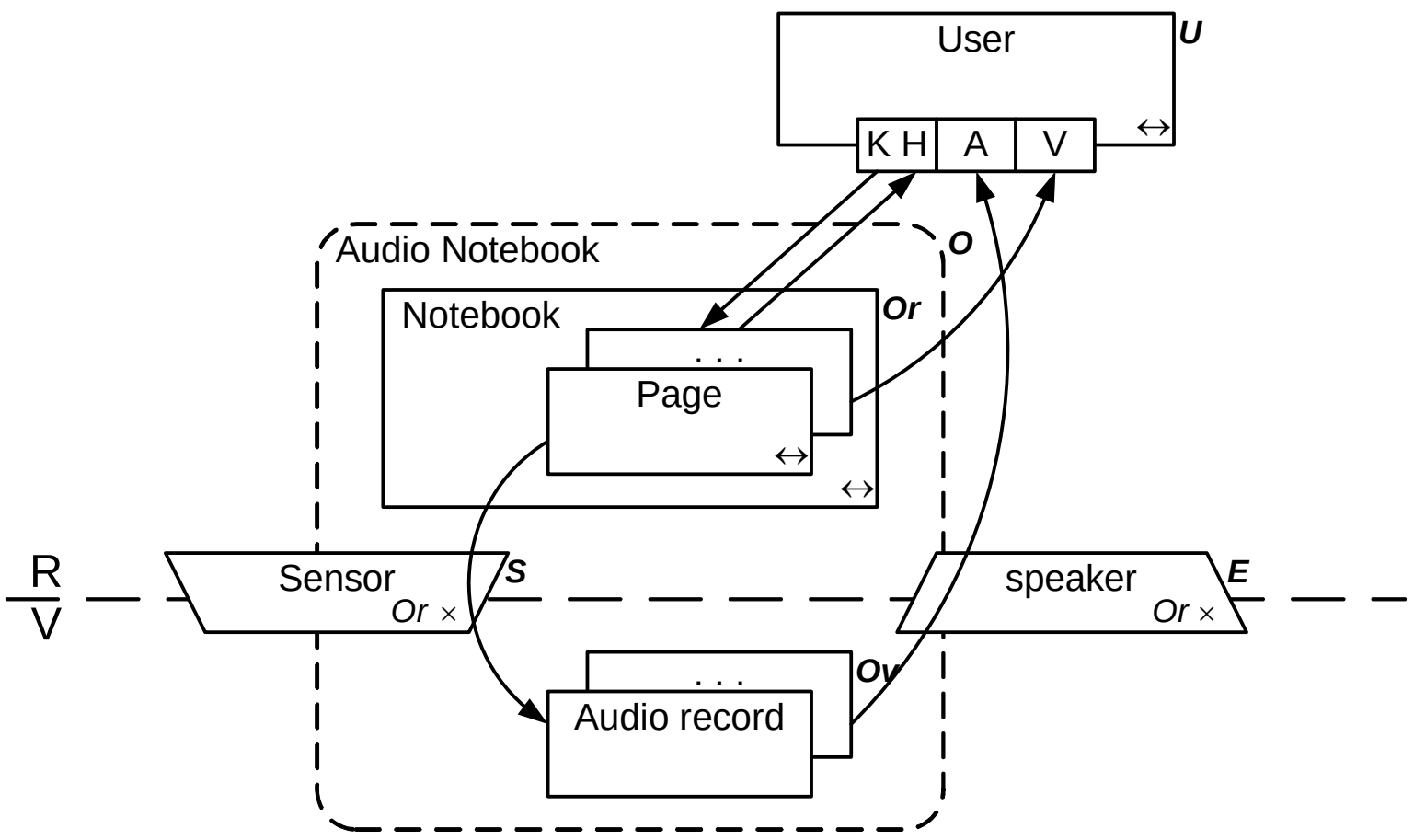

**Figure 6:** IRVO model of Audio Notebook (Stifelman, 1996).

### *3.3.2 DoubleDigitalDesk*

As an example of a collaborative mixed environment, we chose the DoubleDigitalDesk proposed by (Wellner, 1993). In this application two users at a distance draw on a shared sheet of paper thanks to an appropriate set of camera and data projectors. Figure 7 presents the modeling of this application with IRVO. User1 (U1) draws on an ordinary sheet of paper (O1) with a standard pen (T1). Everything s/he’s drawing is captured by a camera (S1) and displayed over the sheet of paper (O2) of User2 (U2) mixed with his/her own drawing thanks to a data projector (E2). This augmented perception is clearly shown by the ⊕ operator of the IRVO notation. Another camera (S2)/data projector (E1) pair captures and overlays User2 drawings back on User1 paper. Globally, the two users have the illusion that they are drawing on the same sheet of paper. As explained in (Wellner, 1993), the signal coming from the camera has to be accurately adapted to limit the infinite loop effect. Nothing in the model deals specifically with this aspect but the graphical notation can help designers to detect this kind of problem. On Figure 7, we also model the audio communication channel between the users composed of two microphones (‘mic.’ sensors S3 and S4) and two speakers (‘spk.’ effectors E3 and E4). Instead of two symmetric arrows between users, a double-ended arrow is used, to avoid cluttering of the model.

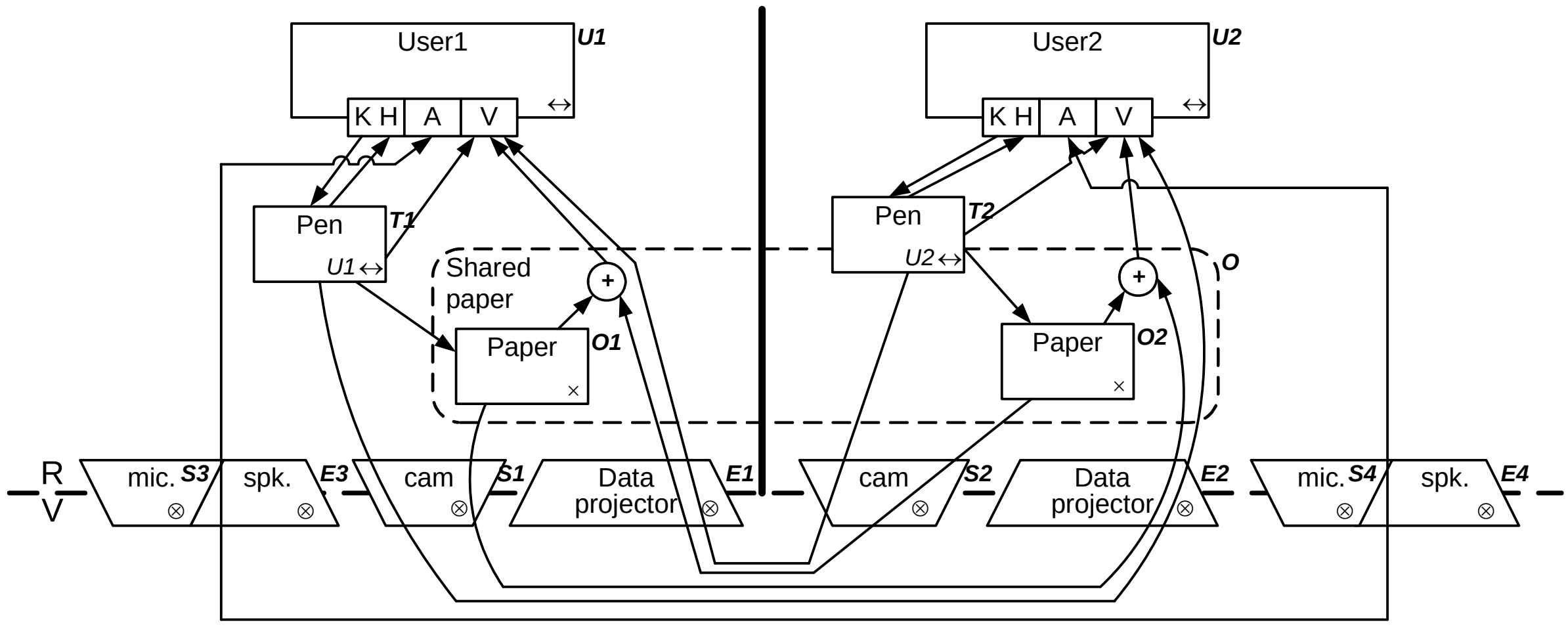

**Figure 7:** IRVO model of DoubleDigitalDesk (Wellner, 1993).

This example presents an interesting case of an augmentation of a real object by another real object. Actually, the second object is not truly 'real' because it is perceived at a distance by the video of the remote scene: (Trevisan, Vanderdonckt & Macq, 2003) call these objects, "digital real objects".

# 4 IRVO use

The above examples (and the 44 modeled applications) show that IRVO has good descriptive abilities. We used this work in (Chalon & David, 2004) to propose a taxonomy of mixed reality systems in a 4x4 matrix helping to classify existing mixed reality applications in sub-categories and even distinguish mixed reality from virtual reality and classical WIMP (Windows, Icons, Menus and Pointing devices) interfaces. These 44 models provide us with a set of typical mixed reality cases which could easily be transformed into a set of interactions patterns suited for designing new applications.

## 4.1 IRVO for designing new applications

The main IRVO objective is to support the design of new applications. Designers use IRVO to model each task of the future application and try different alternatives. To evaluate and compare these different solutions we propose rules that an application must or would follow and which can be checked on the IRVO model of the application:

- An action-perception loop must exist on the model starting from the user (action), going through the tool and the domain objects and going back to the user (perception). If there is no user action in a particular task, then there must only be a perception relation from the domain object to the user (and no tool would appear on the diagram).
- The observability property in HCI says that users must be able to control his/her action. Therefore, there would be a perception relationship on the model between tool and user as well as between domain object and user (which is already taken into account by the previous rule). For mixed objects (either mixed tools or mixed domain objects), each component of the mixed object would be perceived by the user either directly by different senses (as in Audio Notebook) or by merged perception using the ⊕ operator (as in DoubleDigitalDesk)
- Transducers must be used correctly: sensors must be crossed only by relationships coming from the real world and going into the virtual world and effectors must be crossed only by relationships coming from the virtual world and going to the real world. Transducers must be compatible with the intended use: for example a 'screen' can be used to see a virtual object and then is compatible with a relationship going from the virtual object to the 'V' channel of a user.
- The continuity property has been studied by (Dubois et al., 2001). Perceptual continuity is the extension of observability property to the multiple representations (real and virtual) of a single concept. This property can be verified in IRVO models with the ⊕ operator which shows the merging of the perception of several objects as if they were a single object.

- For collaborative applications, the WYSIWIS property says that users must see the same object. This can be verified on an IRVO model if all users perceive the same object. This property is verified in the DoubleDigitalDesk example above.

These rules are generic and can be applied to nearly all situations. Other rules may be added in the future to take into account more specific cases.

## 4.2 Integration of IRVO in a Model-Based Approach

We integrated the IRVO model in CoCSys (Cooperative Capillary Systems), a Model-Based Approach (MBA) which is currently under development in our lab (David, Chalon, Delotte & Vaisman, 2003).
In CoCSys there are two main phases:

- The elaboration of the Collaborative Application Behavioural Model (CAB-M). This model is built by the transformation of scenarios which are sketches of use and a narrative description of an activity within a given context, according to Carroll's definition (Carroll, 1997).
- The Contextualization/Adaptation/Specialization process which transforms and instantiates the CAB-M into a collaborative application based on a 3-level generic framework as per the MDA (Model-Driven Architecture) approach.

In this paper we will focus only on the integration of IRVO in the CAB-M as shown in Figure 8 and the modifications on the first phase of the process. Consequences on the second phase are not included the scope of this paper.

### *4.2.1 Integration in the CAB model*

Before constructing the CAB, scenarios of use of the future applications must be collected either by external observers or by users themselves in a participative design approach. These scenarios are mainly expressed as small narrations (Carroll, 1997) but other forms could be accepted such as UML Use Cases.

The CAB model is elaborated by the transformation of scenarios: the aim is to extract the key elements such as actors, artifacts, tasks, processes and use contexts. This task is mainly carried out manually, but a tool, the CBME editor (Delotte, David & Chalon, 2004), is under development to assist designers in this task. This editor also maintains the link between scenarios and the CAB model and allows complete checking of the CAB. Because scenarios may have superfluous elements, may be incoherent or incomplete, there must be several loop-backs to modify or complete the scenarios until the CAB model becomes coherent and complete.

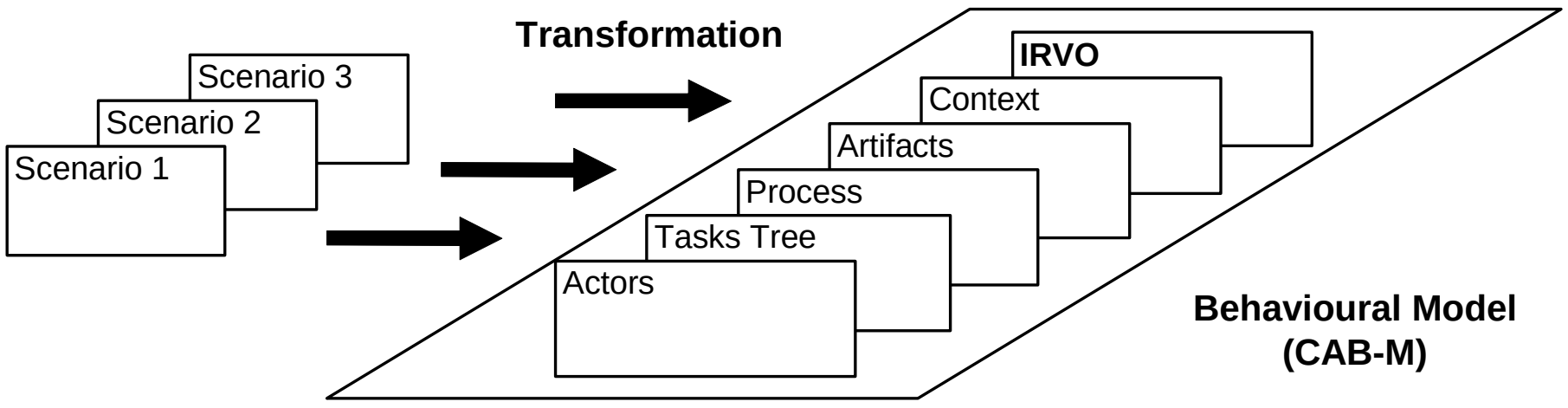

**Figure 8:** Construction of the CAB Model from scenarios.

When building the CAB, some scenarios could exhibit tasks using Mixed Reality. For each of these tasks an IRVO model is created in order to formalize the interaction between users and the system. These models are linked:

- To some tasks (generally leaf tasks) of the tasks tree, as explained in more detail in the following section,
- To the Actors model, in particular by the roles actors are playing in the interaction,
- To the Artifacts model, in particular by the tools that are manipulated and by the domain objects,
- To the Context model, in particular by the devices used.

The IRVO model construction could be based on interaction patterns as explained in section 4.1.
Predictive analysis can be conducted at this stage by evaluating the IRVO models with the rules presented in the previous section and then by comparing alternative solutions. Even if some alternatives do not envisage Mixed Reality, they could be modeled and compared with MR solutions thanks to the capacity of IRVO to model classical WIMP interaction as well as Mixed Reality interaction.

We presented IRVO as a new model that is closely integrated into the CAB model. However, in our opinion it can also be used earlier to support the preparation of scenarios like Renevier's notation (Renevier, 2004). In this context, IRVO can be used to describe graphically specific scenarios where mixed reality is envisaged: these are generally rough diagrams representing main users and objects and omitting 'details' such as tools or transducers at this level. Because IRVO models are graphic models, they could be easily understood by final users who could even sketch them out with the help of designers. These preliminary diagrams can be further enhanced during CAB construction by adding tools and transducers. Because the same notation is used this process is seamless compared to (Renevier, 2004) who proposes changing notation between scenarios (described with his own notation) and the design phase (modeled with ASUR notation).

### *4.2.2 Link with the Tasks tree*

In CoCSys methodology, the Tasks tree uses ConcurTaskTree (CTT) developed by (Paternò, 2000). CTT aims at structuring the tasks in a task tree from basic tasks up to more abstract ones. IRVO aims at modeling the interaction between the user and the artefacts in the context of one task. Therefore it is natural to associate one IRVO diagram with the basic tasks of the task tree (Figure 9a, tasks 11, 12 and 13). If several tasks share the same artifacts, we can directly associate the sub-root of these tasks with the IRVO schema (Figure 9a, task 2).

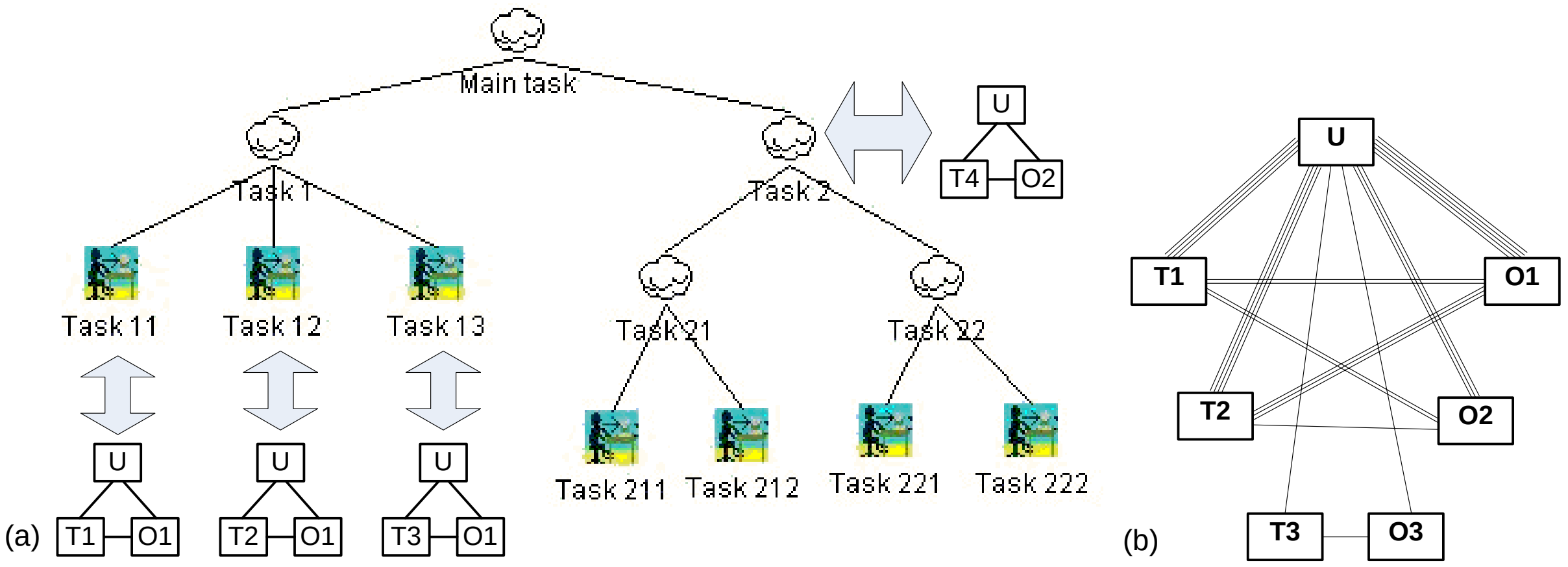

**Figure 9:** links between the Tasks tree and IRVO models.

It appears interesting to return to the root of the task tree by merging IRVO diagrams, which results in a synthetic diagram showing all the artifacts used in the activities (Figure 9b). By analyzing this synthetic diagram we can easily detect odd configurations (such as T3/O3) which have to be examined in order to evaluate their justification (very specific task) or redesigned to comply with the ergonomic rule of interaction continuity.

## 5 Conclusion

In this paper, we presented an interaction model for designing collaborative mixed reality environments. In this model we identified 3 main entities which participate in the interaction: users, objects (tools and domain objects) and the internal model of the application. In the case of mixed reality, only tools and domain objects can be real or virtual. The exchange of data between the real and the virtual worlds is modeled by dedicated entities known as transducers. Two modeling examples of applications taken from the literature demonstrate the descriptive power of this model. We also presented in this paper how IRVO can contribute to a Model-Based Approach, CoCSys, which we are currently designing at our lab. We show that IRVO can be integrated closely into the behavioral model of the application as a new model linked to the other models (Tasks tree and Actors, Artifacts and Context models). In particular, we described in detail the relationship between IRVO models and the Tasks tree.

In this paper we only focused on the IRVO used in the design phase of mixed reality software. We are currently working on the relationships between IRVO and software architecture. In CoCSys methodology, the architectural model used is AMF-C (Multi-Facetted Agents) (Tarpin-Bernard, David & Primet, 1998) and we are examining the consequences of mixed reality on the architectural model of the application. In particular, we propose to extend AMF-C agents by adding a new facet which models the relationship between the presentation layer and the real artifacts.